# Colossal power extraction from active cyclic Brownian information engines

Govind Paneru[a,b,1], Sandipan Dutta[c], and Hyuk Kyu Pak[a,b,1]

[a]Center for Soft and Living Matter, Institute for Basic Science (IBS), Ulsan 44919, Republic of Korea
[b]Department of Physics, Ulsan National Institute of Science and Technology, Ulsan 44919, Republic of Korea
[c]Department of Physics, Birla Institute of Technology and Science, Pilani 333031, India

(Dated: 03-25-2022)

Brownian information engines can extract work from thermal fluctuations by utilizing information. So far, the studies on Brownian information engines consider the system in a thermal bath; however, many processes in nature occur in a nonequilibrium setting, such as the suspensions of self-propelled microorganisms or cellular environments called an active bath. Here, we introduce an archetypal model for Maxwell-demon type cyclic Brownian information engine operating in a Gaussian correlated active bath. The active engine can extract more work than its thermal counterpart, exceeding the bound set by the second law of information thermodynamics. We obtain a general integral fluctuation theorem for the active engine that includes additional mutual information gained from the active bath with a unique effective temperature. This effective description modifies the second law and provides a new upper bound for the extracted work. Unlike the passive information engine operating in a thermal bath, the active information engine extracts colossal power that peaks at the finite cycle period. Our study provides fundamental insights into the design and functioning of synthetic and biological submicron motors in active baths under measurement and feedback control.

Information engines, a modern realization of thought experiments such as Maxwell's demon (1) and the Szilard engine (2), are stochastic devices capable of extracting mechanical work from a single heat bath by exploiting the information acquired from measurements. Recent progress in information thermodynamics has provided the inevitable upper bound of the work that can be extracted from an information engine by generalizing the second law of thermodynamics (3-9):

$$\langle -W \rangle \leq -\Delta F + k_B T \langle \Delta I \rangle, \qquad (1)$$

where $\langle \cdots \rangle$ denotes ensemble average. According to Eq. (1), the average work extracted from an information engine $\langle -W \rangle$ operating in a thermal bath of temperature $T$ is bounded by the associated free energy difference $-\Delta F$ and the average mutual information gain $\langle \Delta I \rangle$ between the system and feedback controller multiplied by $k_B T$, where $k_B$ is the Boltzmann constant.

Various models of information engines operating in thermal baths have been theoretically proposed (5, 6, 10-12) and experimentally verified in classical (13-23) and quantum (24-26) systems. Whether these models and, in particular, the laws of information thermodynamics also apply to information engines operating in athermal baths, such as swimming bacteria and active colloidal particles (27-35) or cellular environments (36-39), remains to be explored.

Brownian particles in such active baths are subject to violent agitation due to the uncorrelated thermal fluctuations of the solvent molecules and the correlated fluctuations generated by the active components. Consequently, they are in a perpetual nonequilibrium state. Recent studies on non-feedback-driven active heat engines operating between active baths of different activity (temperature) reveal that the active heat engines can extract work beyond the limit set by the Carnot bound (40-44). However, due to the limitations in the existing experimental techniques, where it requires a long time to change the activity of the active bath, the active heat engines realized in the experiment operate in the quasistatic limit with cycle time much longer than the thermal relaxation time (40). Moreover, many physiochemical processes in nature occur far from equilibrium in the active bath and exchange energy and information (36, 45, 46). Thus, a more feasible physical model for such processes would be an efficient finite cycle stochastic engine operating in the active bath of constant activity. Here, we introduce an experimentally feasible cyclic information engine operating in an active bath capable of extracting more work than the acquired information, thereby violating the generalized second law in Eq. (1).

The active information engine examined herein consists of a Brownian particle in a harmonic potential well that is subjected to the periodic measurement and feedback control under the influence of Gaussian colored noise, which is a typical model used for active baths (27, 34). We examine the performance of the active information engine as a function of the cycle period, measurement error, and strength and correlation time of the active noise. We find that the thermodynamic quantities such as work, heat, and mutual information of the active engines are greater than those of the passive engines operating in the thermal bath. The average extracted work per cycle in the steady-state where $\Delta F = 0$ can exceed the bound in Eq. (1), but is always bounded by the modified second law $\langle -W \rangle \leq k_B T_{eff} \langle \Delta I \rangle$, where $k_B T_{eff}$ is equivalent to the average effective energy of the particle in the active bath. The modified second law can be also derived from the generalized integral fluctuation theorem that we obtain for the cyclic active information engine as $\langle \exp[-(W/k_B T_{eff} + \Delta I)] \rangle = 1$.

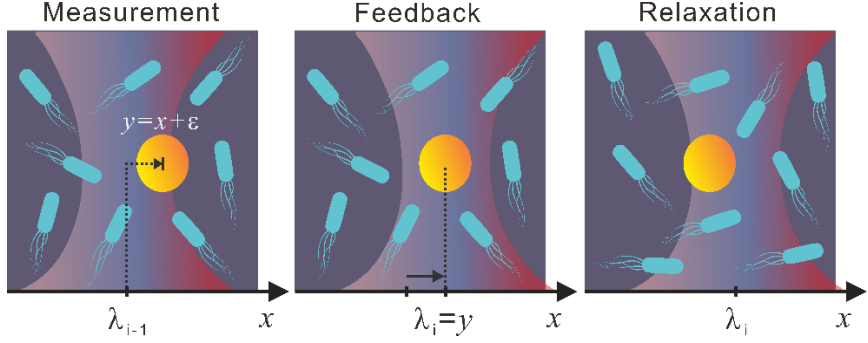

Fig. 1. Illustration of the *i*-th cycle of a Brownian information engine consisting of a colloidal particle in an optical trap operating in an active bath of swimming bacteria. During the measurement step, the information engine perceives the particle position $x$ with respect to the optical trap center $\lambda_{i-1}$ as $y = x + \varepsilon$. The error in the measurement $\varepsilon$ follows a Gaussian white noise of variance $N$. During the feedback step, the trap center is shifted instantaneously to $\lambda_i = y$ and mechanical work is extracted. During the relaxation step, the particle is relaxed in the active bath for a duration $\tau$ with the fixed trap center $\lambda_i$ until the next cycle begins.

One of the key challenges in designing efficient stochastic engines is maximizing the extracted work and power simultaneously (46, 47). We show that the extracted power of the active information engine is maximum for a finite cycle period nearly equal to the thermal relaxation time of the particle where the extracted work is also near maximum. Depending on the active noise parameters, this power can be orders of magnitude larger than those of passive information engines, which exhibit maximum power for ultrafast cycle period where the extracted work vanishes (15, 18, 21). For example, we find that for a strongly correlated active bath of strength $f_{act} \approx 2$ pN and correlation time $\tau_c \approx 25$ ms, the peak power is $\sim 10^4\ k_B T/s$, which is ~50 times larger than its passive counterparts, indicating that the active engines with a finite cycle period can extract a colossal amount of power from the active bath. We also confirm our analytical results using numerical simulations.

*Active bath model–* We consider the one-dimensional motion of a Brownian particle in a harmonic potential, $V(x,\lambda) = (k/2)(x-\lambda)^2$, where $x$ is the particle position, $k$ is the stiffness, and $\lambda$ is the center of the potential in an active bath of temperature $T$. The motion of the particle is described by the overdamped Langevin equation:

$$\gamma \frac{dx}{dt} = -k(x-\lambda) + \xi_{th}(t) + \xi_{act}(t). \quad (2)$$

The thermal noise $\xi_{th}(t)$ follows a Gaussian white noise with zero mean $\langle \xi_{th}(t) \rangle = 0$ and no correlation $\langle \xi_{th}(t)\xi_{th}(t') \rangle = 2\gamma k_B T \delta(t-t')$, where $\gamma$ is the dissipation coefficient. The active noise $\xi_{act}(t)$ is characterized by an exponentially correlated Gaussian noise with a zero mean $\langle \xi_{act}(t) \rangle = 0$, and correlation of (34)

$$\langle \xi_{act}(t)\xi_{act}(t') \rangle = f_{act}^2 \exp(-|t-t'|/\tau_c). \quad (3)$$

Here, $f_{act}$ is the strength and $\tau_c$ is the correlation time of the active noise. In the absence of active noise, the particle is in thermal equilibrium with a Gaussian distribution of $P(x) = (2\pi S)^{-1/2} \exp[-(x-\lambda)^2/2S]$, where $S = k_B T/k$ is the equilibrium variance in the thermal bath. The thermal relaxation time of a particle in the harmonic potential is $\tau_r = \gamma/k$.

In the presence of active noise, the probability distribution function (PDF) of the particle position at any time $t$ still follows a Gaussian distribution but with a variance $S_{act}(t)$, which can be calculated by solving Eq. (2) (SI and (34, 48)):

$$S_{act}(t) = S(0)e^{-2t/\tau_r} + [S + \frac{f_{act}^2 \tau_r}{\gamma^2(1/\tau_r + 1/\tau_c)}][1 - e^{-2t/\tau_r}] \\ - \frac{2f_{act}^2}{\gamma^2(1/\tau_r^2 - 1/\tau_c^2)}[e^{-t(1/\tau_r + 1/\tau_c)} - e^{-2t/\tau_r}], \quad (4)$$

where $S(0)$ is the initial variance of the particle position distribution at $t=0$. Considering the long time limit $t \gg \tau_c$, the active noise correlation decays fully, and the particle reaches a nonequilibrium steady-state. The generalized equipartition theorem can then be defined in the active bath as $\lim_{t \to \infty}(k/2)S_{act} = (k_B/2)(T+T_{act})$ (34), where

$$T_{act} = f_{act}^2/[k_B\gamma(1/\tau_r + 1/\tau_c)] \tag{5}$$

is the active temperature of the particle owing to the active noise source in the medium.

*Active information engine*– Each engine cycle of period $\tau$ includes three steps: particle position measurement, instantaneous shift of the potential center, and relaxation. Figure 1 shows schematics of the *i*th engine cycle operating in the active bath. Here, the information engine measures the particle position $x$ with respect to the potential center $\lambda_{i-1}$ as $y \equiv x + \varepsilon$. The error in the measurement $\varepsilon \equiv y - x$ is characterized by the Gaussian white noise $P(y|x) = (2\pi N)^{-1/2} \exp[-(y-x)^2/2N]$ of variance $N$. During the feedback step, the trap center is shifted instantaneously to the measurement outcome $\lambda_{i-1} \to \lambda_i = y_i$. In the last step, the particle relaxes in the shifted potential before the next cycle begins. The particle dynamics during the relaxation follows the overdamped Langevin Eq. (2). Because the measurement and feedback control are instantaneous, the cycle period $\tau$ is the relaxation period. In the subsequent (*i*+1)th cycle, the origin is reset to the shifted potential center $\lambda_i$, and the same feedback protocol is repeated.

After many repetitions of the feedback cycles, the engine approaches a nonequilibrium steady-state. The PDF of the particle position at the beginning of the relaxation (immediately after the feedback step) is exactly the error distribution $P(y|x)$ with variance $N$. The PDF of the particle position after time $\tau$ (at the beginning of the next cycle) is given by $P(x) = (2\pi S^*(\tau))^{-1/2} \exp[-x^2/2S^*(\tau)]$. The steady-state variance $S^*(\tau)$ is obtained by substituting $S(0) = N$ and $t = \tau$ in Eq. (4).

In the absence of active noise ($f_{act} = 0$), $S^*(\tau)$ reduces to the steady-state variance of a passive information engine operating in a thermal bath of temperature $T$ as $S_{th}^*(\tau) = S + (N-S)e^{-2\tau/\tau_r}$ (21). For ultrafast active and passive engines where $\tau \to 0$, the steady-state variance is equivalent to the variance of the measurement error, $S^*(\tau \to 0) \approx N$. Conversely, the steady-state variance for slower cycle active engines reduces to $S^*(\tau \to \infty) \approx S + f_{act}^2 \tau_r/[\gamma^2(1/\tau_r + 1/\tau_c)]$, which is greater than $S_{th}^*(\tau \to \infty) \approx S$ of the passive engine. For a given cycle period $\tau$, the departure of $S^*(\tau)$ from the thermal equilibrium variance $S$ can be interpreted in terms of the effective temperature of the particle in the active bath under measurement and feedback control:

$$k_B T_{eff}(\tau) = kS^*(\tau). \tag{6}$$

It can be observed from Eqs. (4) to (6) that the effective temperature of the slower cycle active engines is equal to the effective temperature of the particle in the active bath, $T_{eff}(\tau \to \infty) \approx (T+T_{act})$.

Because $P(x)$ and $P(y|x)$ are Gaussian, the PDF of the measurement outcome $P(y) = \int P(x)P(y|x)dx$ is also Gaussian with variance $S^*(\tau) + N$. We can also obtain the conditional PDF immediately after the measurement $P(x|y)$ using Bayes' theorem, $P(x|y)P(y) = P(y|x)P(x)$ (21, 49).

*Thermodynamics of the engine*– The work performed on the particle during each shifting of the potential center is equal to the change in its internal energy plus the heat dissipated into the bath, following the thermodynamic first law (21, 50). However, since the potential is shifted instantaneously after the measurement, the particle has no time to move and dissipate energy. Therefore, the work done on the particle during the feedback step is equal to the change in internal energy $W \equiv \Delta V = (1/2)k[(x-y)^2 - x^2]$. Because the potential center remains fixed during the relaxation step, no work is done on the particle. Hence, the average extracted work by the particle per cycle in the steady-state is given by

$$\langle -W \rangle = -\frac{k}{2}\int dxdy P(y|x)P(x)[(x-y)^2 - x^2]$$
$$= \frac{k}{2}(S^*(\tau) - N). \tag{7}$$

Equation (7) shows the average extracted work $\langle -W \rangle$ is always positive as long as $S^* > N$. The average extracted work per cycle for a passive engine operating in a thermal bath is given by $\langle -W \rangle_{th} = (k/2)(S_{th}^*(\tau) - N)$ (21). Because $S^*(\tau) \geq S_{th}^*(\tau)$, the extracted work for the active engine with a finite cycle period ($\tau > 0$) is always greater than its thermal counterpart, $\langle -W \rangle > \langle -W \rangle_{th}$. The maximum amount of extractable work is $\langle -W \rangle_{max} = (k_B/2)(T + T_{act})$, which is obtained for the error-free active engine ($N=0$) with a slower cycle period ($\tau \to \infty$). Therefore, the error-free and quasistatic cycle active information engines are capable of extracting work equal to the total mean effective energy of the particle in the active bath. The average heat supplied to the system in steady state during the relaxation step is equal to the average extracted work during the feedback, $\langle Q\rangle = \langle -W\rangle$.

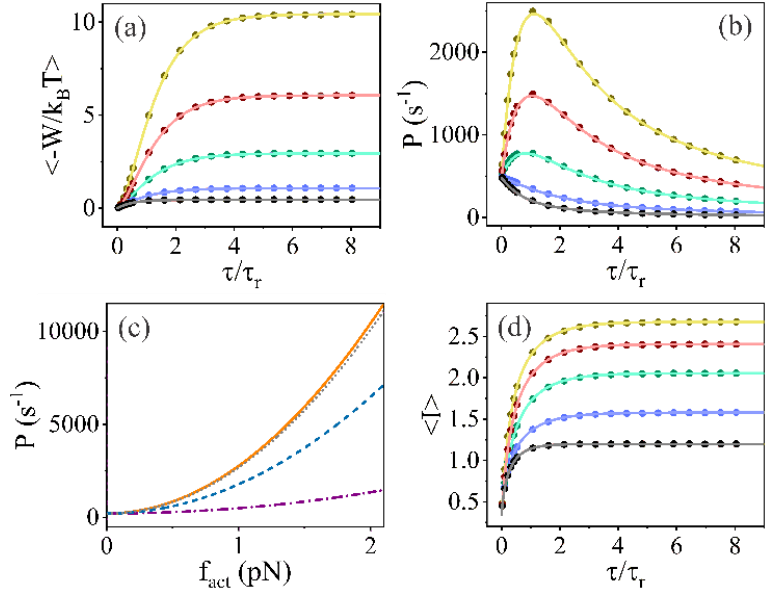

Fig. 2. (a) Average extracted work per cycle $\langle -W/k_B T\rangle$ and (b) average extracted power $P \equiv \langle -W/k_B T\rangle/\tau$ of the information engine as a function of the rescaled cycle period $\tau/\tau_r$ and under a fixed measurement error of $N/S = 0.1$ in a thermal bath (black), as well as in an active bath of fixed correlation time $\tau_c = 25$ ms and noise strength $f_{act} \approx 0.2$ pN (blue), 0.5 pN (olive green), 0.7 pN (burgundy), and 0.9 pN (dark yellow). The solid curves represent the theoretical plots of Eq. (7) in (a), and Eq. (7) divided by $\tau$ in (b). (c) Theoretical plot of $P$ vs $f_{act}$ evaluated at $\tau/\tau_r = 1$ for $\tau_c = 0.1$ ms (purple dashed-dotted), 1 ms (blue dashed), 10 ms (gray dotted), and 25 ms (orange solid). (d) Average mutual information $\langle I\rangle$ per cycle vs $\tau/\tau_r$ under conditions similar to those of panel (a). The solid curves are the theoretical plots of Eq. (8).

We can also find the average mutual information gain for each measurement between the particle position $x$ and the measurement outcome $y$ as

$$\langle I\rangle = \int dxdy P(y|x)P(x)\ln\frac{P(y|x)}{P(y)} = \frac{1}{2}\ln\left(1+\frac{S^*(\tau)}{N}\right). \quad (8)$$

Equation (8) shows that the average mutual information gain by the active engine is greater than that of the passive engine operating in a thermal bath, $\langle I\rangle \geq \langle I\rangle_{th} = (1/2)\ln(1+S^*_{th}(\tau)/N)$.

*Entropy production*– For the information engine, the total entropy production (normalized by $k_B$) per cycle in steady-state is given by $\langle \Delta S_{tot}\rangle = \langle \Delta S_{sys}\rangle + \langle \Delta S_m\rangle + \langle \Delta I\rangle$, where $\Delta S_{sys}$ is the system entropy change, $\Delta S_m$ is the entropy change of the medium, and $\Delta I$ is the net information gain per cycle (23). Since the PDF of the particle position in the steady-state at the beginning of the measurement $P(x,0)$ and at the end of relaxation $P(x,\tau)$ are the same, there is no change in the average system entropy during each cycle, $\langle \Delta S_{sys}\rangle \equiv \langle -\ln P(x,\tau)+\ln P(x,0)\rangle = 0$. In addition, resetting the trap center erases the mutual information between $x$ and $y$, thus $\langle \Delta I\rangle = \langle I\rangle$. The entropy change of the medium can be estimated as the average heat dissipation per cycle in steady-state divided by the effective temperature, $\langle \Delta S_m\rangle \equiv -\langle Q\rangle/k_B T_{eff} = \langle W\rangle/k_B T_{eff}$. Therefore, using the thermodynamic second law $\langle \Delta S_{tot}\rangle = \langle W\rangle/k_B T_{eff} + \langle I\rangle \geq 0$, we obtain the bound for the average extracted work of the cyclic information engine in the active bath:

$$\langle -W\rangle \leq k_B T_{eff}\langle I\rangle. \quad (9)$$

*Integral fluctuation theorem*– We can also modify the generalized integral fluctuation theorem (6), $\langle e^{-(W-\Delta F)/k_B T - \Delta I}\rangle = 1$, for the cyclic information engine operating in an active bath, where $\Delta F = 0$, as (SI)

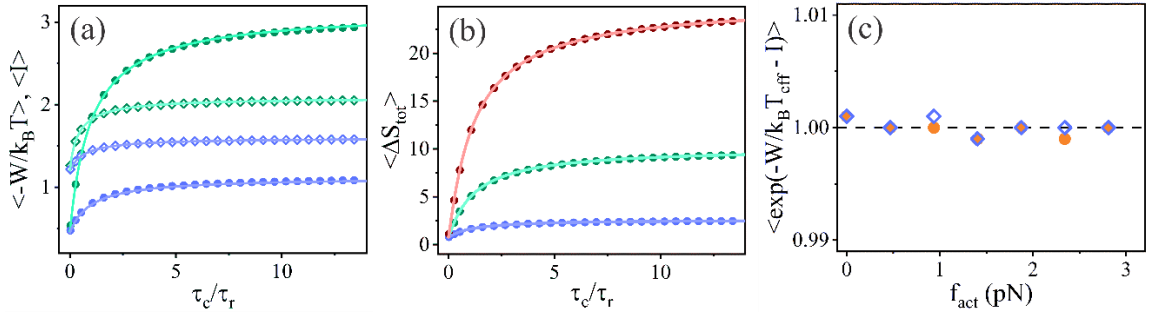

Fig. 3. (a) Plot of the extracted work $\langle -W/k_B T \rangle$ (circles) and mutual information $\langle I \rangle$ (diamonds) per cycle as a function of $\tau_c/\tau_r$ with $f_{act} = 0.2$ pN (blue) and 0.5 pN (olive green) for slower engines $\tau = 15$ ms with a measurement error of $N/S = 0.1$. (b) Average total entropy production per cycle $\langle \Delta S_{tot} \rangle = \langle W \rangle / k_B T_{eff} + \langle I \rangle$ as a function of $\tau_c/\tau_r$ with $f_{act} = 0.2$ pN (blue), 0.5 pN (cyan), and 0.7 pN (burgundy) for $\tau = 15$ ms and $N/S = 0.1$. The solid curves show the theoretical predictions. (c) Plot of $\langle e^{-(W/k_B T_{eff} + I)} \rangle$ as a function of $f_{act}$ for fixed values of $\tau_c = 25$ ms and $N/S = 0.1$ for slower engines $\tau = 15$ ms (orange circles) and faster engines $\tau = 0.5$ ms (blue diamonds).

$$\left\langle e^{-(W/k_B T_{eff} + \Delta I)} \right\rangle = \int dx dy P(y|x) P(x) e^{-(W/k_B T_{eff} + I)} = 1. \quad (10)$$

Note that applying Jansen inequality to Eq. (10) yields the modified generalized second law, which provides a general bound of the extracted work in Eq. (9).

*Results–* We validate the analytical results in Eqs. (7) to (10) via numerical simulations. To achieve this, we numerically solve Eq. (2) for the cyclic information engine using the Euler method with a time step of $\Delta t = 50 \mu s$ and obtain the distributions for the particle position $x$ and the measurement outcome $y$. The input parameters are $T = 293$ K, $\gamma = 6\pi\eta a \approx 18.8$ nNm$^{-1}$s, and $S = (20$ nm$)^2$. The stiffness of the harmonic potential is then $k \equiv k_B T/S \approx 10$ pN/$\mu$m, and the thermal relaxation time of the particle is $\tau_r = \gamma/k \approx 1.88$ ms. We study the performance of the information engine as a function of $f_{act}$ and $\tau_c$ for a fixed distribution of the measurement error $N/S = 0.1$. In addition, we propose that an active information engine with these parameters can be realized in an experiment using the active optical feedback trap technique (21, 51).

Figure 2(a) shows a plot of the average extracted work $\langle -W/k_B T \rangle$ as a function of the rescaled cycle period $\tau/\tau_r$ for a fixed value of $\tau_c \gg \tau_r$ and various values of $f_{act}$. Here, $\langle -W/k_B T \rangle$ is obtained by averaging $-W/k_B T = -(k/2k_B T)[(x-y)^2 - x^2]$ over more than $3.3 \times 10^6$ engine cycles. The numerical results (solid circles) agree well with the theoretical predictions of Eq. (7) (solid curves). We find that, for a given value of $f_{act}$, the extracted work increases with the cycle period and saturates when $\tau \gtrsim 5\tau_r$. For direct comparison, we also plot $\langle -W/k_B T \rangle$ for the passive engine operating in a thermal bath of temperature $T$ (see the black data in Fig. 2(a)). The extracted work for the active engine is always greater than that for the passive engine. The extracted work increases with $f_{act}$, and when $f_{act} \gg f_{th}$, where $f_{th} = \sqrt{k_B T k} \approx 0.2$ pN is the thermal strength of the particle in the harmonic potential, the active engine can extract enormous work from the correlated active bath by exploiting the information about the microstates of the system.

Figure 2(b) shows a plot of the average extracted power $P \equiv \langle -W/k_B T \rangle / \tau$. For the passive information engines, $P$ is maximum for ultrafast engines with a vanishing cycle period $\tau \to 0$ (see the black data in Fig. 2(b)), for which the extracted work vanishes $\langle -W/k_B T \rangle \to 0$. The extracted power $P$ for ultrafast active information engines is equal to that of ultrafast passive engines. However, $P$ for the finite cycle ($\tau > 0$) active information engine is always greater than its passive counterpart. Interestingly, for $f_{act} > f_{th}$ (see the olive green, burgundy, and dark yellow data in Fig. 2(b)), $P$ for the active information engine increases with $\tau$ and reaches a maximum when the cycle period is almost equal to the thermal relaxation time $\tau \approx \tau_r$. We also study the dependence of $P$ on $f_{act}$ and $\tau_c$ for $\tau = \tau_r$ (Fig. 2(c)). For a given $\tau_c > 0$, $P$ increases with $f_{act}$. Whereas for a given $f_{act} > 0$, it increases with $\tau_c$ and saturates when $\tau_c \gtrsim 5\tau_r$. Therefore, the finite-cycle active information engine can extract a colossal amount of power from the active bath by suitably adjusting the active noise parameters.

The peak power observed at the finite cycle period is due to active noise (see Eqs. S9 to S14 and Fig. S4(a)). For $f_{act} \lesssim f_{th}$, thermal noise is dominant and $P$ still exhibits maximum for ultrafast engines $\tau \to 0$. We find that $P$ peaks at a finite cycle period ($\tau > 0$) only when $f_{act} \gtrsim f_{th}$ and $\tau_c > 0$. The peak position shifts towards the higher values of $\tau$ on increasing $f_{act}$ or $\tau_c$, and saturates to $\tau \approx 1.26 \tau_r$ when $f_{act} >> f_{th}$ and $\tau_c >> \tau_r$ (Fig. S4(b)).

Figure 2(d) shows the average mutual information $\langle I \rangle$ between $x$ and $y$ as a function of $\tau/\tau_r$ under similar conditions as in Fig. 2(a) ($\tau_c >> \tau_r$ and varied $f_{act}$). Here, $\langle I \rangle$ is obtained by averaging $\ln[P(y|x)/P(y)]$. The numerical results (solid circles) agree well with the theoretical predictions of Eq. (8) (solid curves). We find that $\langle I \rangle$ increases with $\tau$ and saturates for slower engines $\tau >> \tau_r$. The saturated value of $\langle I \rangle$ is greater than that of the passive engine. In addition, $\langle I \rangle$ increases with the strength of the active noise.

Next, we study the performance of the active engine as a function of $\tau_c$ for different values of $f_{act}$. The extracted work $\langle -W/k_B T \rangle$ and mutual information $\langle I \rangle$ increase with $\tau_c$ and saturate when $\tau_c >> \tau_r$. For passive engines, the extracted work is always bounded by the mutual information, $\langle -W/k_B T \rangle \leq \langle I \rangle$, following the generalized second law in Eq. (1). For active engines, this is true only for relatively smaller active noise strengths, and when $f_{act} >> f_{th}$, the saturated value of $\langle -W/k_B T \rangle$ can exceed $\langle I \rangle$ (see the olive green data in Fig. 3(a) and Fig. S2). However, it is shown that $\langle -W \rangle$ is always bounded by $k_B T_{eff} \langle I \rangle$. To further examine this phenomenon, we measure the total entropy production per cycle $\langle \Delta S_{tot} \rangle = \langle W \rangle / k_B T_{eff} + \langle I \rangle$ as a function of $\tau_c$, as shown in Fig. 3(b). $\langle \Delta S_{tot} \rangle$ increases with $\tau_c$ and saturates when $\tau_c >> \tau_r$. Moreover, we find that $\langle \Delta S_{tot} \rangle > 0$, thereby validating the modified generalized second law in Eq. (9) (Fig. 3(b) and Fig. S3). Finally, we test the integral fluctuation theorem in Eq. (10). To achieve this, we evaluate $\langle e^{-(W/k_B T_{eff} + I)} \rangle$ as a function of $f_{act}$ for $\tau_c >> \tau_r$ and find it to be equal to unity irrespective of the cycle period (Fig. 3(c)).

In conclusion, we extended the generalized integral fluctuation theorem and the second law of thermodynamics towards the situations in which the system extracts work from an active bath by exploiting positional information concerning the state of the system. We studied the performance of a cyclic Brownian information engine in the presence of Gaussian colored noise. The engine can extract the maximum work equal to the mean effective energy of the active bath, thereby exceeding the conventional bound of the generalized second law of thermodynamics. However, the extracted work is bounded by the new upper bound, which includes an additional information gain due to the active noise source in the solution. Unlike passive information engines, the extracted power of the active information engines is maximum at a finite cycle period near the thermal relaxation time of the particle. This study provides fundamental insights into the manipulating of energy and information in nonequilibrium systems under fluctuating and correlated environments.

**Acknowledgements**
We thank Cheol-Min Ghim for critically reading the manuscript and providing fruitful comments. This research was supported by the Institute for Basic Science (grant no. IBS-R020-D1).

[1]To whom correspondence should be addressed. Email: gpaneru@gmail.com or hyuk.k.pak@gmail.com

# Supplementary Material

**Variance of the probability distribution function of the particle position in the active bath**

The solution of the overdamped Langevin equation in Eq. (2) in the main text can be obtained by using Laplace's transform method:

$$x(t) = x(0)e^{-t/\tau_r} + \frac{1}{\gamma}\int_0^t dt' e^{-(t-t')/\tau_r}[\xi_{th}(t') + \xi_{act}(t')]. \tag{S1}$$

Using the fact that the averages of the thermal and active noises are zero, $\langle \xi_{th}(t) \rangle = 0$ and $\langle \xi_{act}(t) \rangle = 0$, and their correlations are $\langle \xi_{th}(t)\xi_{th}(t') \rangle = 2\gamma k_B T \delta(t-t')$ and $\langle \xi_{act}(t)\xi_{act}(t') \rangle = f_{act}^2 e^{-|t-t'|/\tau_c}$, we get the variance of the probability distribution function of the particle in the active bath, $S_{eff}(t) = \langle x^2(t) \rangle$, at any time $t > 0$ as

$$\begin{aligned}\langle x^2(t) \rangle &= \langle x^2(0) \rangle e^{-2t/\tau_r} + \frac{1}{\gamma^2}\int_0^t dt' e^{-(t-t')/\tau_r}\int_0^t dt'' e^{-(t-t'')/\tau_r}[\langle \xi_{th}(t')\xi_{th}(t'') \rangle + \langle \xi_{act}(t')\xi_{act}(t'') \rangle] \\ &= \langle x^2(0) \rangle e^{-2t/\tau_r} + \frac{1}{\gamma^2}\int_0^t dt' e^{-(t-t')/\tau_r}\int_0^t dt'' e^{-(t-t'')/\tau_r}[2\gamma k_B T \delta(t'-t'') + f_{act}^2 e^{-|t'-t''|/\tau_c}].\end{aligned} \tag{S2}$$

Equation (S2) can be simplified to Eq. (4) in the main text as

$$S_{act}(t) = S(0)e^{-2t/\tau_r} + [S + \frac{f_{act}^2 \tau_r}{\gamma^2(1/\tau_r + 1/\tau_c)}][1 - e^{-2t/\tau_r}] - \frac{2f_{act}^2}{\gamma^2(1/\tau_r^2 - 1/\tau_c^2)}[e^{-t(1/\tau_r + 1/\tau_c)} - e^{-2t/\tau_r}], \tag{S3}$$

where $S(0) = \langle x^2(0) \rangle$ is the initial variance at $t = 0$.

**Steady-state variance for active information engine**

The engine configuration for a given cycle period $\tau$ is specified by the particle position $x$ and the position of the trap center $\lambda$. During the feedback step, the potential center is shifted to the measurement outcome $y$. In a relative frame of reference, the trap center is fixed at the origin while the particle position is instantaneously reset to $-y$. Thus, the probability distribution of the particle position in steady-state immediately after the feedback is given by the measurement error distribution $P(y|x) = (2\pi N)^{-1/2} \exp[-(y-x)^2/2N]$ [1]. On substituting $S(0) = N$ in Eq. (S3), we get the value for $S^*(\tau)$ as

$$S^*(\tau) = N e^{-2\tau/\tau_r} + [S + \frac{f_{act}^2 \tau_r}{\gamma^2(1/\tau_r + 1/\tau_c)}][1 - e^{-2\tau/\tau_r}] - \frac{2f_{act}^2}{\gamma^2(1/\tau_r^2 - 1/\tau_c^2)}[e^{-\tau(1/\tau_r + 1/\tau_c)} - e^{-2\tau/\tau_r}]. \tag{S4}$$

We define the effective temperature of the active engine in a steady state as

$$T_{eff}(\tau) = (kS^*(\tau))/k_B. \tag{S5}$$

The plot of $T_{eff}(\tau)$ as a function of cycle period $\tau$ and active noise correlation time $\tau_c$ for the cyclic active engine in steady state is shown in Fig. S1.

**Modified generalized integral fluctuation theorem**

The generalized integral fluctuation $\langle e^{-(W-\Delta F)/k_B T - \Delta I} \rangle = 1$ [2,3] is modified for the cyclic active engine in steady state, where $\Delta F = 0$, by taking account of the effective temperature $T_{eff}(\tau)$ as

$$\langle e^{-[(W-\Delta F)/k_B T_{eff} + \Delta I]} \rangle = \int dx dy P(y|x) P(x) e^{-(W/k_B T_{eff} + I)}. \tag{S6}$$

Using $I = \ln[P(y|x)/P(y)]$ we show that the modified generalized integral fluctuation theorem is equal to one as

$$\langle e^{-[(W-\Delta F)/k_B T_{eff} + \Delta I]} \rangle = \int dx dy P(x) P(y) e^{-((x-y)^2 - x^2)/2S^*} = 1. \tag{S7}$$

Using Jansen's inequality, we get the modified generalized second law (Eq. (9) in the main text).

$$\langle W \rangle / k_B T_{eff} + \langle \Delta I \rangle \geq 0. \tag{S8}$$

**Maximum power**

The average extracted power in steady state is given by

$$P = \langle -W \rangle / \tau = \frac{k}{2\tau} (S^*(\tau) - N). \tag{S9}$$

Equation (S9) can be expressed as a sum of power due to thermal bath $P_{th}$ and active bath $P_{act}$ as

$$P = P_{th} + P_{act}, \tag{S10}$$

where

$$P_{th} = \frac{k}{2\tau}(S - N)(1 - e^{-2\tau/\tau_r}), \tag{S11}$$

and

$$P_{act} = \frac{k}{2\tau} \left[ \frac{f_{act}^2 \tau_r}{\gamma^2 (1/\tau_r + 1/\tau_c)} (1 - e^{-2\tau/\tau_r}) - \frac{2 f_{act}^2}{\gamma^2 (1/\tau_r^2 - 1/\tau_c^2)} (e^{-\tau(1/\tau_r + 1/\tau_c)} - e^{-2\tau/\tau_r}) \right]. \tag{S12}$$

For slower cycle engines $\tau \to \infty$, both $P_{th}$ and $P_{act}$ vanishes, therefore, $P(\tau \to \infty) \approx 0$. For ultrafast engines $\tau \to 0$, the dependence of $P$ on $P_{th}$ and $P_{act}$ can be obtained by doing Taylor series expansion of the exponential functions up to the third term:

$$P_{th}(\tau \to 0) \approx \frac{(f_{th}^2 - f_{error}^2)}{\gamma} \left(1 - \frac{\tau}{\tau_r}\right) \approx \frac{(f_{th}^2 - f_{error}^2)}{\gamma}, \tag{S13}$$

and

$$P_{act}(\tau \to 0) \approx \frac{\tau f_{act}^2}{2\tau_r \gamma} \approx 0, \tag{S14}$$

where $f_{th} = \sqrt{k^2 S}$ and $f_{error} = \sqrt{k^2 N}$. Thus, the average power for ultrafast engines $\tau \to 0$ is equal to that of the passive engines, $P(\tau \to 0) \approx (f_{th}^2 - f_{error}^2)/\gamma$.

The observed peak power at the finite cycle period of the active information engine is due to the presence of active noise (Fig. S4(a)). As shown above, $P_{act}$ vanishes for both ultrafast and slower cycle engines. It assumes a maximum at a finite value of $\tau$. The peak position in $P_{act}$ depends on $\tau_c$. It shifts to the higher value of $\tau$ and saturates to $\tau \approx 1.26\tau_r$ for $\tau_c \to \infty$.

For $f_{act} < f_{th}$, thermal noise is dominant, and $P$ is maximum for ultrafast cycle periods $\tau \to 0$. $P_{act}$ is dominant only when $f_{act} > f_{th}$ and $\tau_c > 0$, consequently $P$ assumes a maximum at a finite $\tau$. The peak position in $P$ is controlled by $f_{act}$ and $\tau_c$ of the active noise. It shifts to a higher value of $\tau$ with increase in $f_{act}$ or $\tau_c$, and saturates to $\tau \approx 1.26\tau_r$ for $f_{act} \gg f_{th}$ and $\tau_c \gg \tau_r$ (Fig. S4 (b)).

**Figures**

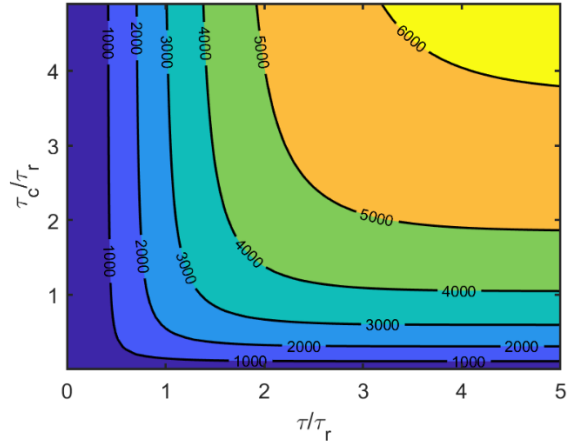

Fig. S1. Plot of the effective temperature $T_{eff}(\tau)$ in Kelvin (Eq. S5) for the active engine in steady state as a function of cycle period $\tau/\tau_r$ and correlation time $\tau_c/\tau_r$ for the fixed value of the active noise strength $f_{act} = 1$ pN.

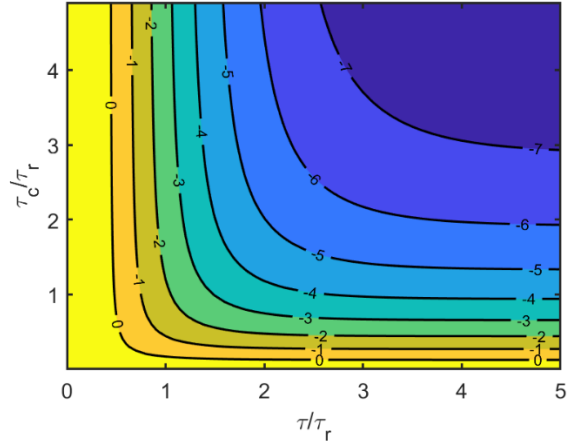

Fig. S2. Violation of the generalized Second law, $\langle W \rangle / k_B T + \langle \Delta I \rangle \geq 0$. Theoretical plot of $\langle W \rangle / k_B T + \langle \Delta I \rangle$ for the active engine in steady state as a function of cycle period $\tau/\tau_r$ and correlation time $\tau_c/\tau_r$ for the fixed value of the active noise strength $f_{act} = 1$ pN.

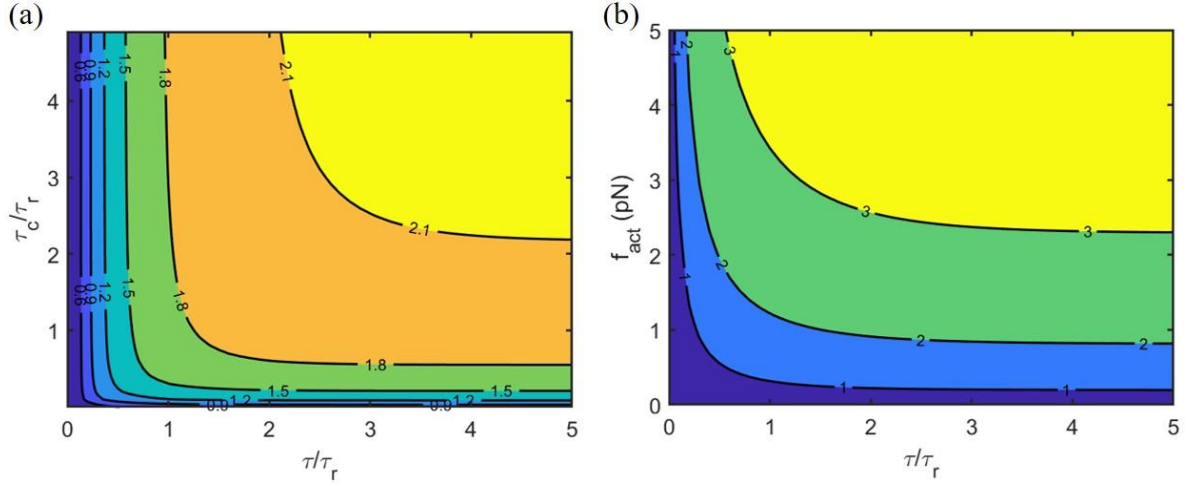

Fig. S3. Modified generalized second law, $\langle W \rangle / k_B T_{eff} + \langle \Delta I \rangle \geq 0$, is valid. (a) Theoretical plot of $\langle W \rangle / k_B T_{eff} + \langle \Delta I \rangle$ for the active engine in steady state as a function of cycle period $\tau / \tau_r$ and correlation time $\tau_c / \tau_r$ for the fixed value of the active noise strength $f_{act} = 1$ pN. (b) Plot of $\langle W \rangle / k_B T_{eff} + \langle \Delta I \rangle$ as a function of as a function of $\tau / \tau_r$ and $f_{act}$ for the fixed value of $\tau_c / \tau_r = 5$.

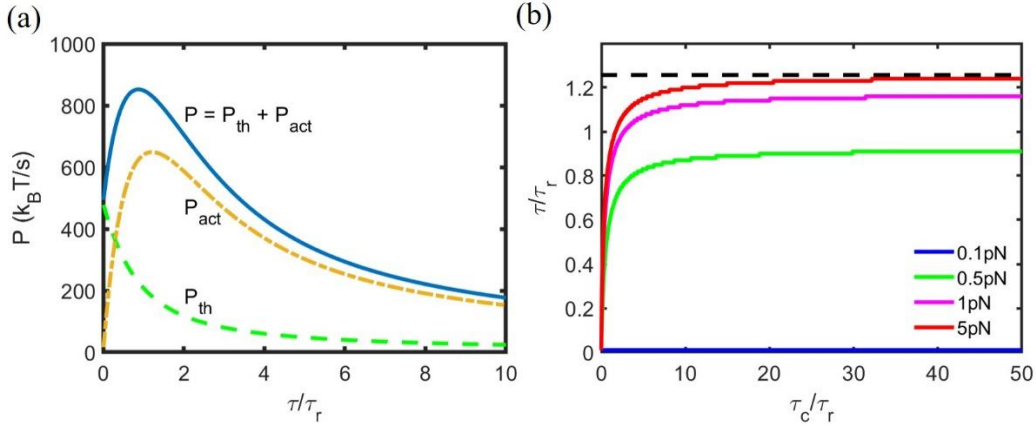

Fig. S4. (a) Plot of average extracted power in Eq. (S9) (blue), power due to thermal bath in Eq. (S11) (green), and power due to active bath in Eq. (S12) (orange) as a function of $\tau / \tau_r$ for fixed $\tau_c / \tau_r \approx 13$ and $f_{act} / f_{th} \approx 2.5$. (b) Plot showing the peak position of the extracted power shifts towards the higher value of $\tau$ with increase in $f_{act}$ and $\tau_c$ and saturates to $\tau \approx 1.26 \tau_r$ when $f_{act} \gg f_{th}$ and $\tau_c \gg \tau_r$.